\catcode`\@=11					



\font\fiverm=cmr5				
\font\fivemi=cmmi5				
\font\fivesy=cmsy5				
\font\fivebf=cmbx5				

\skewchar\fivemi='177
\skewchar\fivesy='60


\font\sixrm=cmr6				
\font\sixi=cmmi6				
\font\sixsy=cmsy6				
\font\sixbf=cmbx6				

\skewchar\sixi='177
\skewchar\sixsy='60


\font\sevenrm=cmr7				
\font\seveni=cmmi7				
\font\sevensy=cmsy7				
\font\sevenit=cmti7				
\font\sevenbf=cmbx7				

\skewchar\seveni='177
\skewchar\sevensy='60


\font\eightrm=cmr8				
\font\eighti=cmmi8				
\font\eightsy=cmsy8				
\font\eightit=cmti8				
\font\eightbf=cmbx8				

\skewchar\eighti='177
\skewchar\eightsy='60


\font\ninei=cmmi9
\font\ninesy=cmsy9

\skewchar\ninei='177
\skewchar\ninesy='60


\font\tenrm=cmr10				
\font\teni=cmmi10				
\font\tensy=cmsy10				
\font\tenex=cmex10				
\font\tenit=cmti10				
\font\tensl=cmsl10				
\font\tenbf=cmbx10				
\font\tentt=cmtt10				
\font\tenss=cmss10				
\font\tensc=cmcsc10				
\font\tenbi=cmmib10				

\skewchar\teni='177
\skewchar\tenbi='177
\skewchar\tensy='60

\def\tenpoint{\ifmmode\err@badsizechange\else
	\textfont0=\tenrm \scriptfont0=\sevenrm \scriptscriptfont0=\fiverm
	\textfont1=\teni  \scriptfont1=\seveni  \scriptscriptfont1=\fivemi
	\textfont2=\tensy \scriptfont2=\sevensy \scriptscriptfont2=\fivesy
	\textfont3=\tenex \scriptfont3=\tenex   \scriptscriptfont3=\tenex
	\textfont4=\tenit \scriptfont4=\sevenit \scriptscriptfont4=\sevenit
	\textfont5=\tensl
	\textfont6=\tenbf \scriptfont6=\sevenbf \scriptscriptfont6=\fivebf
	\textfont7=\tentt
	\textfont8=\tenbi \scriptfont8=\seveni  \scriptscriptfont8=\fivemi
	\def\rm{\tenrm\fam=0 }%
	\def\it{\tenit\fam=4 }%
	\def\sl{\tensl\fam=5 }%
	\def\bf{\tenbf\fam=6 }%
	\def\tt{\tentt\fam=7 }%
	\def\ss{\tenss}%
	\def\sc{\tensc}%
	\def\bmit{\fam=8 }%
	\rm\setparameters\setbaselines\fi}


\font\twelverm=cmr12				
\font\twelvei=cmmi12				
\font\twelvesy=cmsy10	scaled\magstep1		
\font\twelveex=cmex10	scaled\magstep1		
\font\twelveit=cmti12				
\font\twelvesl=cmsl12				
\font\twelvebf=cmbx12				
\font\twelvett=cmtt12				
\font\twelvess=cmss12				
\font\twelvesc=cmcsc10	scaled\magstep1		
\font\twelvebi=cmmib10	scaled\magstep1		

\skewchar\twelvei='177
\skewchar\twelvebi='177
\skewchar\twelvesy='60

\def\twelvepoint{\ifmmode\err@badsizechange\else
	\textfont0=\twelverm \scriptfont0=\eightrm \scriptscriptfont0=\sixrm
	\textfont1=\twelvei  \scriptfont1=\eighti  \scriptscriptfont1=\sixi
	\textfont2=\twelvesy \scriptfont2=\eightsy \scriptscriptfont2=\sixsy
	\textfont3=\twelveex \scriptfont3=\tenex   \scriptscriptfont3=\tenex
	\textfont4=\twelveit \scriptfont4=\eightit \scriptscriptfont4=\sevenit
	\textfont5=\twelvesl
	\textfont6=\twelvebf \scriptfont6=\eightbf \scriptscriptfont6=\sixbf
	\textfont7=\twelvett
	\textfont8=\twelvebi \scriptfont8=\eighti  \scriptscriptfont8=\sixi
	\def\rm{\twelverm\fam=0 }%
	\def\it{\twelveit\fam=4 }%
	\def\sl{\twelvesl\fam=5 }%
	\def\bf{\twelvebf\fam=6 }%
	\def\tt{\twelvett\fam=7 }%
	\def\ss{\twelvess}%
	\def\sc{\twelvesc}%
	\def\bmit{\fam=8 }%
	\rm\setparameters\setbaselines\fi}


\font\fourteenrm=cmr12	scaled\magstep1		
\font\fourteeni=cmmi12	scaled\magstep1		
\font\fourteensy=cmsy10	scaled\magstep2		
\font\fourteenex=cmex10	scaled\magstep2		
\font\fourteenit=cmti12	scaled\magstep1		
\font\fourteensl=cmsl12	scaled\magstep1		
\font\fourteenbf=cmbx12	scaled\magstep1		
\font\fourteentt=cmtt12	scaled\magstep1		
\font\fourteenss=cmss12	scaled\magstep1		
\font\fourteensc=cmcsc10 scaled\magstep2	
\font\fourteenbi=cmmib10 scaled\magstep2	

\skewchar\fourteeni='177
\skewchar\fourteenbi='177
\skewchar\fourteensy='60

\def\fourteenpoint{\ifmmode\err@badsizechange\else
	\textfont0=\fourteenrm \scriptfont0=\tenrm \scriptscriptfont0=\sevenrm
	\textfont1=\fourteeni  \scriptfont1=\teni  \scriptscriptfont1=\seveni
	\textfont2=\fourteensy \scriptfont2=\tensy \scriptscriptfont2=\sevensy
	\textfont3=\fourteenex \scriptfont3=\tenex \scriptscriptfont3=\tenex
	\textfont4=\fourteenit \scriptfont4=\tenit \scriptscriptfont4=\sevenit
	\textfont5=\fourteensl
	\textfont6=\fourteenbf \scriptfont6=\tenbf \scriptscriptfont6=\sevenbf
	\textfont7=\fourteentt
	\textfont8=\fourteenbi \scriptfont8=\tenbi \scriptscriptfont8=\seveni
	\def\rm{\fourteenrm\fam=0 }%
	\def\it{\fourteenit\fam=4 }%
	\def\sl{\fourteensl\fam=5 }%
	\def\bf{\fourteenbf\fam=6 }%
	\def\tt{\fourteentt\fam=7}%
	\def\ss{\fourteenss}%
	\def\sc{\fourteensc}%
	\def\bmit{\fam=8 }%
	\rm\setparameters\setbaselines\fi}


\font\seventeenrm=cmr10 scaled\magstep3		


\newdimen\rp@
\newcount\@basestretchnum
\newskip\@baseskip
\newskip\headskip
\newskip\footskip


\def\setparameters{\rp@=.1em
	\headskip=24\rp@
	\footskip=\headskip
	\delimitershortfall=5\rp@
	\nulldelimiterspace=1.2\rp@
	\scriptspace=0.5\rp@
	\abovedisplayskip=10\rp@ plus3\rp@ minus5\rp@
	\belowdisplayskip=10\rp@ plus3\rp@ minus5\rp@
	\abovedisplayshortskip=5\rp@ plus2\rp@ minus4\rp@
	\belowdisplayshortskip=10\rp@ plus3\rp@ minus5\rp@
	\normallineskip=\rp@
	\lineskip=\normallineskip
	\normallineskiplimit=0pt
	\lineskiplimit=\normallineskiplimit
	\jot=3\rp@
	\setbox0=\hbox{\the\textfont3 B}\p@renwd=\wd0
	\skip\footins=12\rp@ plus3\rp@ minus3\rp@
	\skip\topins=0pt plus0pt minus0pt}


\def\setbaselines{\maxdepth=4\rp@\baselinestretch=\@basestretchnum}


\def\baselinestretch{\afterassignment\@basestretch\@basestretchnum}
\def\@basestretch{%
	\@baseskip=12\rp@ \divide\@baseskip by1000
	\normalbaselineskip=\@basestretchnum\@baseskip
	\baselineskip=\normalbaselineskip
	\bigskipamount=\the\baselineskip
		plus.25\baselineskip minus.25\baselineskip
	\medskipamount=.5\baselineskip
		plus.125\baselineskip minus.125\baselineskip
	\smallskipamount=.25\baselineskip
		plus.0625\baselineskip minus.0625\baselineskip
	\setbox\strutbox=\hbox{\vrule height.708\baselineskip
		depth.292\baselineskip width0pt }}



\def\makeheadline{\vbox to0pt{\baselinestretch=1000
	\vskip-\headskip \vskip1.5pt
	\line{\vbox to\ht\strutbox{}\the\headline}\vss}\nointerlineskip}

\def\makefootline{\baselineskip=\footskip\line{\the\footline}}

\def\big#1{{\hbox{$\left#1\vbox to8.5\rp@ {}\right.\n@space$}}}
\def\Big#1{{\hbox{$\left#1\vbox to11.5\rp@ {}\right.\n@space$}}}
\def\bigg#1{{\hbox{$\left#1\vbox to14.5\rp@ {}\right.\n@space$}}}
\def\Bigg#1{{\hbox{$\left#1\vbox to17.5\rp@ {}\right.\n@space$}}}


\mathchardef\alpha="710B
\mathchardef\beta="710C
\mathchardef\gamma="710D
\mathchardef\delta="710E
\mathchardef\epsilon="710F
\mathchardef\zeta="7110
\mathchardef\eta="7111
\mathchardef\theta="7112
\mathchardef\iota="7113
\mathchardef\kappa="7114
\mathchardef\lambda="7115
\mathchardef\mu="7116
\mathchardef\nu="7117
\mathchardef\xi="7118
\mathchardef\pi="7119
\mathchardef\rho="711A
\mathchardef\sigma="711B
\mathchardef\tau="711C
\mathchardef\upsilon="711D
\mathchardef\phi="711E
\mathchardef\chi="711F
\mathchardef\psi="7120
\mathchardef\omega="7121
\mathchardef\varepsilon="7122
\mathchardef\vartheta="7123
\mathchardef\varpi="7124
\mathchardef\varrho="7125
\mathchardef\varsigma="7126
\mathchardef\varphi="7127
\mathchardef\imath="717B
\mathchardef\jmath="717C
\mathchardef\ell="7160
\mathchardef\wp="717D
\mathchardef\partial="7140
\mathchardef\flat="715B
\mathchardef\natural="715C
\mathchardef\sharp="715D


\def\err@badsizechange{%
	\immediate\write16{--> Size change not allowed in math mode, ignored}}

\baselinestretch=1000
\tenpoint

\catcode`\@=12					
\catcode`\@=11
\expandafter\ifx\csname @iasmacros\endcsname\relax
	\global\let\@iasmacros=\par
\else	\immediate\write16{}
	\immediate\write16{Warning:}
	\immediate\write16{You have tried to input iasmacros more than once.}
	\immediate\write16{}
	\endinput
\fi
\catcode`\@=12


\def\rmb{\seventeenrm}

\def\singlespace{\baselineskip=\normalbaselineskip}
\def\halfspace{\baselineskip=1.5\normalbaselineskip}
\def\doublespace{\baselineskip=2\normalbaselineskip}


\def\AB{\bigskip\parindent=40pt
        \centerline{\bf ABSTRACT}\medskip\halfspace\narrower}
\def\AE{\bigskip\nonarrower\doublespace}
\def\nonarrower{\advance\leftskip by-\parindent
	\advance\rightskip by-\parindent}


\def\boxit#1{\vbox{\hrule\hbox{\vrule\kern3pt
	\vbox{\kern3pt#1\kern3pt}\kern3pt\vrule}\hrule}}

\def\hence{\leavevmode\hbox{\bf .\raise5.5pt\hbox{.}.} }

\def\dalemb#1#2{{\vbox{\hrule height.#2pt
	\hbox{\vrule width.#2pt height#1pt \kern#1pt \vrule width.#2pt}
	\hrule height.#2pt}}}
\def\gtorder{\mathrel{\raise.3ex\hbox{$>$}\mkern-14mu
             \lower0.6ex\hbox{$\sim$}}}
\def\ltorder{\mathrel{\raise.3ex\hbox{$<$}\mkern-14mu
             \lower0.6ex\hbox{$\sim$}}}

\newdimen\fullhsize
\newbox\leftcolumn
\def\twoup{\hoffset=-.5in \voffset=-.25in
  \hsize=4.75in \fullhsize=10in \vsize=6.9in
  \def\fullline{\hbox to\fullhsize}
  \let\lr=L
  \output={\if L\lr
        \global\setbox\leftcolumn=\columnbox\global\let\lr=R \advancepageno
      \else \doubleformat \global\let\lr=L\fi
    \ifnum\outputpenalty>-20000 \else\dosupereject\fi}
  \def\doubleformat{\shipout\vbox{
    \fullline{\box\leftcolumn\hfil\columnbox}\advancepageno}}
  \def\columnbox{\leftline{\vbox{\makeheadline\pagebody\makefootline}}}
  \tolerance=1000 }
\twelvepoint
\doublespace
{\nopagenumbers{
\rightline{~~~December, 2000}
\bigskip\bigskip
\centerline{\rmb 
Symmetry Breaking for Matter Coupled to Linearized Supergravity}   
\centerline{\rmb 
>From the Perspective of the Current Supermultiplet}

\medskip
\centerline{\it Stephen L. Adler
}
\centerline{\bf Institute for Advanced Study}
\centerline{\bf Princeton, NJ 08540}
\medskip
{\singlespace{
\leftline{{\it 4th Award in the 2001 Gravity Research Foundation Essay 
Competition}} 
\leftline{{\it To be published in General Relativity and Gravitation}}
 }}
\bigskip\bigskip
\leftline{\it Send correspondence to:}
\medskip
{\singlespace\leftline{Stephen L. Adler}
\leftline{Institute for Advanced Study}
\leftline{Einstein Drive, Princeton, NJ 08540}
\leftline{Phone 609-734-8051; FAX 609-924-8399; 
email adler@ias.edu}}
\bigskip\bigskip
}}
\vfill\eject
\pageno=2
\AB
We consider a generic supersymmetric matter theory coupled to 
linearized supergravity, and analyze scenarios for spontaneous 
symmetry breaking in terms of vacuum expectation values of components 
of the current supermultiplet.  When the vacuum expectation of the 
energy momentum tensor is zero, but the scalar current or pseudoscalar 
current gets an expectation, evaluation of the gravitino self energy  
using the supersymmetry current algebra shows that there is an induced 
gravitino mass term.   The structure of this term  generalizes the  
supergravity action with cosmological constant to theories with $CP$   
violation.  When the vacuum expectation of the energy momentum 
tensor is nonzero, supersymmetry is broken;  requiring cancellation of 
the cosmological constant gives the corresponding generalized  
gravitino mass formula.   
\AE
\bigskip\bigskip
\vfill\eject
\pageno=3

Supersymmetry, to be relevant to physics, must be broken, and 
mechanisms for supersymmetry breaking have been intensively studied.   
In this essay, we shall analyze scenarios for spontaneous symmetry 
breaking in locally supersymmetric theories by reference to 
the vacuum expectation values of the components  
of the current supermultiplet, through which a generic supersymmetric 
matter theory couples to linearized supergravity.   

In linearized general relativity, the spacetime metric $g_{\mu\nu}$ 
deviates from the Minkowski metric $\eta_{\mu\nu}$
by a small perturbation $h_{\mu\nu}$,  
$$g_{\mu\nu}=\eta_{\mu\nu}+2\kappa h_{\mu\nu}~~~,\eqno(1)$$
with the proportionality  constant $\kappa$ related to Newton's   
constant $G$ and the Planck mass $M_{\rm Planck}$ by  
$$\kappa=(8\pi G)^{1\over 2}=M_{\rm Planck}^{-1}~~~.\eqno(2)$$
In linearized supergravity, one adjoins to the spin 2 graviton field 
$h_{\mu\nu}$ 
a spin $3/2$ Rarita-Schwinger Majorana field $\psi_{\mu}$, which 
describes the fermionic gravitino partner of the graviton.  
A gravity supermultiplet, for which the supersymmetry 
algebra closes without use of the equations of motion, is obtained 
by adding auxiliary fields, consisting [1] of an axial 
vector $b_{\mu}$, a scalar $M$, and a 
pseudoscalar $N$.  The supersymmetry variations which close the 
supersymmetry algebra 
(with constant Grassmann supersymmetry parameter $\epsilon$, and with 
$a \cdot c \equiv a_{\mu}c^{\mu}$) are  
$$\eqalign{
\delta h_{\mu\nu}=&{1\over 2} \overline{\epsilon}(\gamma_{\mu}\psi_{\nu}
+\gamma_{\nu}\psi_{\mu})~~~,\cr
\delta \psi_{\mu}=&[-\sigma^{\kappa\nu} \partial_{\kappa} h_{\nu\mu}
-{1\over 3} \gamma_{\mu}(M+i\gamma_5 N) +(b_{\mu}-{1\over 3} \gamma_{\mu} 
\gamma \cdot b) i \gamma_5] \epsilon~~~,\cr
\delta b_{\mu}=&{3\over 2} i \overline{\epsilon} \gamma_5 (R_{\mu}-
{1\over 3} \gamma_{\mu} \gamma \cdot R)~,~~~~
\delta M=-{1\over 2} \overline{\epsilon} \gamma \cdot R~,~~~~
\delta N=-{1\over 2} i \overline{\epsilon} \gamma_5 \gamma \cdot R~~~.\cr
}\eqno(3)$$
The corresponding linearized supergravity action, which  
is invariant under these variations, is
$$S_{\rm grav}=\int d^4x[E^{\mu\nu}h_{\mu\nu}-{1\over 2}\overline{\psi}_{\mu}
R^{\mu}-{1\over 3}(M^2+N^2-b_{\mu}b^{\mu})]~~~,\eqno(4)$$
with $E^{\mu\nu}$ the linearized Einstein tensor and with  
$R^{\nu}=i\epsilon^{\nu\mu\kappa\rho}\gamma_5\gamma_{\mu}\partial_{\kappa}
\psi_{\rho}.$   

Linearized supergravity couples to supersymmetric matter through a real
supermultiplet of currents [2], consisting of the energy momentum tensor 
$\theta^{\mu\nu}$, the supersymmetry current $j_{\mu}$, an axial 
vector current $j_{\mu}^{(5)}$, a scalar 
density $P$, and a pseudoscalar density $Q$.  These transform [3] under 
supersymmetry variations as 
$$\eqalign{
\delta \theta^{\mu\nu}=&{1\over 4}\overline{\epsilon}
(\sigma^{\kappa\mu}\partial_{\kappa}j^{\nu}+\sigma^{\kappa\nu}
\partial_{\kappa}j^{\mu})~~~,\cr
\delta j_{\mu}=&[2\gamma^{\nu}\theta_{\mu\nu}-i\gamma_5\gamma\cdot\partial
j_{\mu}^{(5)}+i\gamma_5\gamma_{\mu}\partial \cdot j^{(5)}
+{1\over 2}\epsilon_{\mu\nu\rho\kappa}\gamma^{\nu}\partial^{\rho}
j^{\kappa (5)} +{1 \over 3}\sigma_{\mu\nu}\partial^{\nu}(P+i\gamma_5 Q)]
\epsilon~~~,\cr
\delta j_{\mu}^{(5)}=&i\overline{\epsilon}\gamma_5 j_{\mu}
-{1\over 3}i \overline{\epsilon}\gamma_5 \gamma_{\mu} \gamma \cdot j~,~~~~
\delta P=\overline{\epsilon} \gamma \cdot j~,~~~~
\delta Q=i\overline{\epsilon}\gamma_5 \gamma \cdot j~~~.\cr
}\eqno(5)$$
The matter interaction action that 
is invariant under simultaneous supersymmetry variations of 
the gravity and current supermultiplets, and that gives the correct  
Newtonian static limit, is  
$$S_{\rm int}=\kappa \int d^4x  [h_{\mu\nu}\theta^{\mu\nu}
+{1 \over 2} \overline{\psi}_{\mu}j^{\mu}-{1\over 2} b_{\mu}j^{\mu(5)}
-{1\over 6}(MP+NQ)]~~~. \eqno(6)$$
Since the auxiliary 
fields $b_{\mu}$, $M$, and $N$ enter  
with no differential operators acting on them, their equations of motion 
following from Eqs.~(4) and (6)
are the algebraic relations 
$$b_{\mu}={3\over 4}\kappa j_{\mu}^{(5)}~,~~~~ 
M=-{1\over 4} \kappa P~,~~~~
N=-{1\over 4} \kappa Q~~~.\eqno(7)$$

Using Eq.~(7), one can eliminate the auxiliary fields  from the 
combined supergravity and interaction actions.   As we have recently 
shown [3], by completing the square one can also eliminate 
the graviton and gravitino fields from the linearized theory.  
This gives the full effective 
action $S_{\rm eff}$ which describes the order $\kappa^2$ back reaction of 
supergravity on the matter sector, 
$$\eqalign{
S_{\rm eff}=&\kappa^2 \int d^4x \left[-{3\over 16} j_{\mu}^{(5)}
j^{\mu(5)} +{1\over 48} (P^2+Q^2)\right] \cr
+&\kappa^2 \int d^4xd^4y\left[{1\over 4}\theta^{\nu\tau}(x)
(\eta_{\nu\alpha}\eta_{\tau\beta}+\eta_{\nu\beta}\eta_{\tau\alpha}
-\eta_{\nu\tau}\eta_{\alpha\beta})\Delta_F(x-y)\theta^{\alpha\beta}(y)
\right.\cr
-&\left. {1\over 8} \overline{j}_{\tau}(x)\left(\eta^{\tau\nu}\gamma 
\cdot \partial_x +{1\over 2}\gamma^{\tau} \gamma \cdot \partial_x 
\gamma^{\nu}\right) \Delta_F(x-y) j_{\nu}(y)\right]~~~,\cr}\eqno(8)$$
with $\Delta_F$ the massless Feynman propagator   
$$\Delta_F(x-y)={1\over (2\pi)^4}\int d^4q {e^{iq \cdot (x-y)}  
\over q^2-i0^+}~~~.\eqno(9)$$
Using conservation of the 
currents $j_{\mu}$ and $\theta_{\mu\nu}$, one can show that Eq.~(8) is 
invariant under the supersymmetry transformation on the current 
supermultiplet given in Eq.~(5).

To examine the implications of the above relations 
for spontaneous symmetry breaking, we take vacuum expectations 
(denoted by 
$\langle~~~ \rangle$) of Eqs. (5) and (7).   
Because Lorentz  
invariance requires the vanishing of the vacuum expectations  
$\langle j_{\mu}^{(5)}\rangle $,  $\langle j_{\mu} \rangle$, and $\langle  
b_{\mu} \rangle$, while $\langle \theta_{\mu\nu}\rangle $ can be proportional 
to the Minkowski metric $\eta_{\mu\nu}$, and so can be nonzero, 
Eq.~(5) gives 
$$\eqalign{
\langle \delta \theta^{\mu\nu}\rangle=&\langle \delta j_{\mu}^{(5)}\rangle=
\langle \delta P\rangle =\langle \delta Q\rangle =0~~~,\cr
\langle \delta j_{\mu}\rangle=&[2\gamma^{\nu}\langle \theta_{\mu\nu} \rangle
+{1 \over 3}\sigma_{\mu\nu}\partial^{\nu}(\langle P\rangle +i\gamma_5 
\langle Q\rangle )]
\epsilon~~~,\cr
}\eqno(10)$$
and Eq.~(7) gives  
$$\langle M\rangle=-{1\over 4} \kappa \langle P\rangle~,~~
\langle N\rangle =-{1\over 4} \kappa \langle Q\rangle~~~.\eqno(11)$$
Since $\langle P\rangle$, $\langle Q \rangle$ are 
coordinate independent by translation invariance,  
they do not contribute to the right hand side of Eq.~(10).  Hence Eq.~(10) 
for $\langle \delta j_{\mu}\rangle$ simplifies to 
$$\langle \delta j_{\mu}\rangle=2\gamma^{\nu}
\langle \theta_{\mu\nu} \rangle
\epsilon~~~.\eqno(12)$$

Let us first consider the case when $\langle \theta_{\mu\nu}\rangle=0$.
The set of expectations 
$$\eqalign{
\langle \theta_{\mu\nu}\rangle=&\langle j_{\mu}\rangle =
\langle j_{\mu}^{(5)}\rangle=0~~~,\cr
\langle P \rangle\not=&0~,~~~~\langle Q \rangle\not=0~~~,\cr
}\eqno(13)$$
satisfy Eqs.~(10) if we take the supersymmetry variations of the   
expectations to be $\delta \langle P \rangle = \langle \delta P \rangle =0$,  
$\delta \langle Q \rangle=\langle \delta Q \rangle =0$.  
Thus, the transformation properties of the supermultiplet 
of currents are preserved  
when the scalar current $P$ and the pseudoscalar current $Q$ develop 
nonzero vacuum expectations that are supersymmetry invariants.  

Whether $P$ and/or $Q$ have nonzero expectations is  
a matter of detailed dynamics. 
An important case where $\langle P\rangle \not=0$, but supersymmetry  
remains unbroken, is 
supersymmetric Yang-Mills theory.  In this theory $P$ is related to the  
gaugino density through the scalar component of the anomaly supermultiplet, 
$$P=g^{-1}\beta(g) \overline{\chi}\chi ~~~,
\eqno(14)$$  
with $g$ the Yang-Mills coupling, and hence $P$ develops a nonzero 
expectation, 
$$\langle P \rangle = g^{-1} \beta(g) 
\langle \overline{\chi} \chi \rangle~~~,\eqno(15)$$
as a result of the formation [4] of a vacuum gaugino condensate.  

When $P$ and/or $Q$ has a nonzero expectation, Eq.~(8) implies a nonzero  
vacuum energy density (the negative of the vacuum action density)   
given by 
$$\rho_{\rm VAC}=-{\kappa^2 \over 48} ( \langle P \rangle^2  +
\langle Q \rangle^2)
~~~.\eqno(16)$$
There are two other places where effects arising from  $\langle P \rangle$ 
and $\langle Q \rangle$ 
appear.  First, from Eqs.~(3) and (11), we see that the supersymmetry 
variation of the gravitino field receives a contribution from 
$\langle P \rangle$ and $\langle Q \rangle$ given by 
$$\delta \psi_{\mu} = {\kappa \over 12} \gamma_{\mu} 
(\langle P \rangle +i \gamma_5 \langle Q 
\rangle)\epsilon +...~~~, \eqno(17)$$
with $...$ denoting terms with expectation zero.  

Second, $\langle P \rangle$ and $\langle Q \rangle$ contribute to the 
gravitino self energy.  
To order $\kappa^2$, the gravitino self energy 
induced by matter couplings is
given by the action addition
$$\Delta S=i {\kappa^2 \over 8} \int d^4x d^4y \overline{\psi}_{\mu A}
(x) \langle T(j_A ^{\mu}(x)\overline{j}_B^{\rho}(y))\rangle 
\psi_{\rho B}(y)~~~,\eqno(18)$$
with $A,B$ spinor indices.  The action term involving no 
derivatives of the gravitino field is obtained by treating the gravitino 
field as a constant in Eq.~(18), leading to 
$$\Delta S\simeq i {\kappa^2 \over 8} 
\int d^4x \overline{\psi}_{\mu A}
(x) \langle K_{AB}^{\mu\rho} \rangle \psi_{\rho B}(x)~~~,\eqno(19)$$
with the constant operator $K_{AB}^{\mu\rho}$ defined by 
$$K_{AB}^{\mu\rho}\equiv{ \int d^4x \int d^4y  
T(j_A ^{\mu}(x)\overline{j}_B^{\rho}(y))
\over \int d^4x~1}
~~~.\eqno(20)$$ 
To evaluate $K_{AB}^{\mu\rho}$, we use current algebra methods, 
by expanding the identity 
$$0=\int d^4x d^4y {\partial \over x^{\theta}} [x^{\mu}
T(j_A ^{\theta}(x)\overline{j}_B^{\rho}(y)) ]~~~,\eqno(21)$$
giving 
$$K_{AB}^{\mu\rho}={-\int d^4xd^4y  x^{\mu}[T(\partial \cdot j_A(x)
\overline{j}_B^{\rho}(y))
+\delta(x^0-y^0)\{j_A^0(x),\overline{j}_B^{\rho}(y)\}]
\over \int d^4x ~1}~~~.\eqno(22)$$
Using conservation of $j_A^{\theta}$, together with the fact that $j_A^0$ 
is the supersymmetry generator obeying 
$$\overline{\epsilon}_A \{j_A^0(x),\overline{j}_B^{\rho}(y)\}=
-i\delta^3(\vec x-\vec y)\delta \overline{j}_B^{\rho}(y)~~~,\eqno(23)$$ 
and using Eq.~(5) to calculate the supersymmetry 
variation on the right hand side of Eq.~(23), we get 
$$\eqalign{
K_{AB}^{\mu\rho}=&{-\int d^4xd^4y x^{\mu} \delta^4(x-y) 
\left({-i\over 3}\right)
\sigma_{AB}^{\tau \rho} {\partial \over y^{\tau}}[P(y)+i\gamma_5 Q(y)]+...
\over \int d^4x ~1 } \cr
=&-{i\over 3} \sigma^{\mu\rho}_{AB}  {\int d^4x[P(x)+i\gamma_5 Q(x)] +...
\over \int d^4x~1 }~~~.\cr
}\eqno(24)$$
The terms denoted by $...$ do not contribute to the expectation, and so 
Eq.~(24) implies 
$$\langle K_{AB}^{\mu\rho}\rangle=-{i\over 3} \sigma^{\mu\rho}_{AB} 
(\langle P\rangle+i\gamma_5\langle Q \rangle) ~~~,\eqno(25)$$
which when substituted into Eq.~(19) gives the gravitino mass term 
$$\Delta S_{\rm mass} = {\kappa^2 \over 24}  
\int d^4x \overline{\psi}_{\mu}(x)
(\langle P \rangle+i \gamma_5\langle Q \rangle)  
\sigma^{\mu\rho}  \psi_{\rho}(x)~~.\eqno(26)$$
When the CP violating expectation $\langle Q\rangle$ is zero, 
Eqs.~(16), 
(17), and (26) are respectively the vacuum energy density, the modified 
gravitino variation, 
and the gravitino mass term that enter into the extension [5] of 
supergravity to accommodate 
a nonvanishing cosmological constant, corresponding to supergravity in   
anti-de Sitter space [6].  When the expectation $\langle Q 
\rangle$ is nonzero, these equations give a generalized supergravity 
with cosmological constant, in which there is also a $CP$ violating  
gravitino mass term.  

This generalized supergravity is supersymmetric even beyond the linearized 
approximation.  To see this, we make the polar decomposition   
$$\langle P \rangle + i \gamma_5 \langle Q \rangle
=(\langle P \rangle^2 + \langle Q \rangle^2)^{1\over 2} 
e^{i\alpha \gamma_5}~,~~~~ 
\alpha=\arctan(\langle Q \rangle / \langle P \rangle)~~~.\eqno(27)$$ 
and define new gamma matrices $\tilde{\gamma}_{\mu}=\gamma_{\mu}\exp(i
\alpha \gamma_5)=\exp(-i\alpha\gamma_5/2) 
\gamma_{\mu}\exp(i\alpha\gamma_5/2)$, which obey the same identities as  
the $\gamma_{\mu}$, as well as $\tilde{\gamma}_{\mu}\tilde{\gamma}_{\nu}
=\gamma_{\mu}\gamma_{\nu}$.  Since  
$\overline{\psi}$ contains a factor $\gamma^0$, we see that 
Eqs.~(17) and (26) plus the gravitino kinetic term are equivalent to  
the theory with $\langle Q \rangle =0$, with $\langle P \rangle $ replaced 
by $(\langle P \rangle^2 + \langle Q \rangle^2)^{1\over 2}$, and with all  
$\gamma_{\mu}$ replaced by the corresponding $\tilde{\gamma}_{\mu}$, 
to which the supersymmetry proofs of Refs. [5] apply.  

Let us turn now to the  generic case, in which the expectations 
$\langle \theta_{\mu\nu} \rangle$, 
$\langle P \rangle$, and $\langle Q \rangle$ are all nonzero.  Using  
$\langle \theta_{\mu\nu}\rangle = \langle \theta_0^0 \rangle \eta_{\mu\nu}$, 
Eq.~(12) can be rewritten as 
$$\langle \delta j_{\mu} \rangle = 2 \gamma_{\mu} \langle \theta_0^0 \rangle 
\epsilon~~~.\eqno(28)$$
Since a nonzero value of $\langle j_{\mu} \rangle $ implies 
that supersymmetry 
is broken, we recover the usual criterion, 
that supersymmetry in the matter sector is broken if and only if the 
positive semidefinite matter 
vacuum energy density $\langle \theta_0^0 \rangle$ is nonzero.  Adding  
$\langle \theta_0^0 \rangle$ to Eq.~(16), the total vacuum energy 
density becomes  
$$\rho_{\rm VAC}= \langle \theta_0^0\rangle 
-{\kappa^2 \over 48} (\langle P \rangle^2 + \langle Q \rangle^2)~~~.
\eqno(29)$$
Rewriting Eq.~(26) as 
$$\eqalign{
\Delta S_{\rm mass}=&
{1\over 2} m
\int d^4x \overline{\psi}_{\mu}  (x) \sigma^{\mu\rho}  \psi_{\rho}(x)
+{1\over 2} m^{\prime}
\int d^4x \overline{\psi}_{\mu}  (x)i 
\gamma_5 \sigma^{\mu\rho} \psi_{\rho}(x)
~~~,\cr
m=&{\kappa^2 \over 12} \langle P \rangle~,~~~~
m^{\prime}={\kappa^2 \over 12} \langle Q \rangle~~~,\cr
}\eqno(30)$$
the condition for the vacuum energy density 
$\rho_{\rm VAC}$ of Eq.~(29) to vanish by cancellation 
between the matter and supergravity contributions is  
$$\kappa \left[ \langle \theta_0^0 \rangle \over 3 \right]^{1\over 2} = 
{\kappa^2 \over 12}(\langle P \rangle^2 + \langle Q \rangle^2)^{1\over 2}=
(m^2 + m^{\prime~2})^{1\over 2}~~~.\eqno(31)$$
We thus obtain a new derivation (when $\langle Q \rangle$=$m^{\prime}=0$) 
of the Deser-Zumino [7] formula for the gravitino 
mass, as well as  its extension to the case when the 
$CP$ violating expectation $\langle Q \rangle$ is nonzero.

To conclude, we have shown that by using the transformation  
properties of the current supermultiplet,  one can analyze     
possibilities for supersymmetry breaking when supersymmetric matter is  
coupled to linearized supergravity.  Nonlinear supergravity corrections 
to our results appear only at higher orders in the expansion in powers of 
$\kappa$. In addition to giving a compact current-algebraic derivation of 
the action for supergravity with a cosmological constant, and of the 
gravitino mass formula, our method generalizes  these 
results to the case when the matter theory breaks 
$CP$ invariance, allowing the expectation $\langle Q \rangle$ to be nonzero.
 
\vfill 
\eject

\bigskip
\centerline{\bf Acknowledgments}
This work was supported in part by the Department of Energy under
Grant \#DE--FG02--90ER40542.  The author wishes to thank  
D. Boulware, C. Burgess, A. Kapustin, J. Maldacena, G. Moore, 
A. Nelson, N. Seiberg, E. Witten, and L. Yaffe for helpful comments.    
\bigskip

\bigskip

\centerline{\bf References}
\bigskip
\noindent
[1]  Stelle, K., and West, P. (1978). 
{\it Phys. Lett. B}{\bf 74}, 330; Ferrara, S., 
and van Nieuwenhuizen, P. (1978). 
{\it Phys. Lett. B}{\bf 74}, 333.
\hfill\break
\bigskip
\noindent
[2] Ferrara, S., and Zumino, B. (1975).  {\it Nucl. Phys. B}{\bf 87}, 207.
\hfill\break
\bigskip
\noindent
[3]  These formulas are taken from  Adler, S. L. (2000), ``Completing the 
Square to Find the Supersymmetric Matter Effective Action Induced 
by Coupling to Linearized $N=1$ Supergravity'', hep-th/0009069, 
submitted to {\it Annals of Physics}.  
\hfill\break
\bigskip
\noindent
[4]  Nilles, H. P. (1982). {\it Phys. Lett. B}{\bf 112}, 455;  
Veneziano, G., and 
Yankielowicz, S. (1982). {\it Phys. Lett. B}{\bf 113}, 231;  Davis, A. C.,  
Dine, M., and Seiberg, N. (1983). {\it Phys. Lett. B}{\bf 125}, 487; 
Affleck, I., Dine, M., and Seiberg,N. (1984).  
{\it Nucl. Phys. B}{\bf 241}, 493.
\hfill\break
\bigskip
\noindent
[5]  Freedman, D. Z., and Das, A. (1977). {\it Nucl Phys. B}{\bf 120}, 221; 
MacDowell, S. W.,  and Mansouri, F. (1977).  {\it Phys. Rev. Lett.} 
{\bf 38}, 739; 
Townsend, P. K. (1977). {\it Phys. Rev. D}{\bf 15}, 2802.   Our formulation 
is closest to that of Townsend.  Note that 
$\surd 2 \kappa$ in Townsend's paper is our $\kappa$, and $\epsilon$ 
of Townsend's paper is our $\surd 2 \epsilon$, so that $\kappa \epsilon$ is 
the same in both.  
\hfill\break
\bigskip
\noindent
[6] I wish to thank E. Witten for alerting me to the role of   
anti-de Sitter supergravity in this discussion.  
\hfill\break
\bigskip
\noindent
[7]  Deser, S., and  Zumino, B. (1977). {\it Phys. Rev. Lett.} 
{\bf 38}, 1433.  
 For an alternative derivation 
(which omits the $m^{\prime}$ mass term of Eq.~(30) and so is valid only 
when $\langle Q \rangle =0$) see Weinberg, S. (2000) {\it The Quantum 
Theory of Fields, Volume III Supersymmetry} (Cambridge University Press, 
Cambridge), Secs. 29.2 and 31.3.  
\hfill\break
\bigskip 
\noindent
\vfill
\eject
\bigskip
\bye